\documentclass[conference]{IEEEtran}
\IEEEoverridecommandlockouts
\usepackage{cite}
\usepackage{amsmath,amssymb,amsfonts}
\usepackage{algorithmic}
\usepackage{hyperref}
\usepackage{cleveref}
\usepackage{graphicx}
\usepackage{caption}
\usepackage{subcaption}
\usepackage{amsmath}
\usepackage{adjustbox}
\usepackage{tcolorbox}
\usepackage{textcomp}
\usepackage{xcolor}
\usepackage{hyperref}
\usepackage{multirow}
\usepackage{booktabs}
\def\BibTeX{{\rm B\kern-.05em{\sc i\kern-.025em b}\kern-.08em
    T\kern-.1667em\lower.7ex\hbox{E}\kern-.125emX}}
    
\begin{document}

\title{Evaluation of Distributed Data Processing Frameworks in Hybrid Clouds\\
}

\author{\IEEEauthorblockN{Faheem Ullah, Shagun Dhingra, Xiaoyu Xia, M.Ali Babar}
\IEEEauthorblockA{
% \textit{CREST\textsuperscript{1}\thanks{\noindent\textsuperscript{1}Centre for Research on Engineering Software Technologies}, The University of Adelaide}\\
School of Computer Science, The University of Adelaide, Australia\\
\{faheem.ullah, shagun.dhingra, xiaoyu.xia, ali.babar\}@adelaide.edu.au}
% \and
% \IEEEauthorblockN{Shagun Dhingra}
% \IEEEauthorblockA{\textit{School of Computer Science} \\
% \textit{CREST, The University of Adelaide}\\
% Adelaide, Australia\\
% shagun.dhingra@adelaide.edu.au}
% \and
% \IEEEauthorblockN{M. Ali Babar}
% \IEEEauthorblockA{\textit{School of Computer Science} \\
% \textit{CREST, The University of Adelaide}\\
% Adelaide, Australia\\
% ali.babar@adelaide.edu.au}
}
\maketitle
\thispagestyle{plain}
\pagestyle{plain}
\begin{abstract}
Distributed data processing frameworks (e.g., Hadoop, Spark, and Flink) are widely used to distribute data among computing nodes of a cloud. Recently, there have been increasing efforts aimed at evaluating the performance of distributed data processing frameworks hosted in private and public clouds. However, there is a paucity of research on evaluating the performance of these frameworks hosted in a hybrid cloud, which is an emerging cloud model that integrates private and public clouds to use the best of both worlds. Therefore, in this paper, we evaluate the performance of Hadoop, Spark, and Flink in a hybrid cloud in terms of execution time, resource utilization, horizontal scalability, vertical scalability, and cost. For this study, our hybrid cloud consists of OpenStack (private cloud) and MS Azure (public cloud). We use both batch and iterative workloads for the evaluation. Our results show that in a hybrid cloud (i) the execution time increases as more nodes are borrowed by the private cloud from the public cloud, (ii) Flink outperforms Spark, which in turn outperforms Hadoop in terms of execution time, (iii) Hadoop transfers the largest amount of data among the nodes during the workload execution while Spark transfers the least amount of data, (iv) all three frameworks horizontally scale better as compared to vertical scaling, and (v) Spark is found to be least expensive in terms of \$ cost for data processing while Hadoop is found the most expensive.
\end{abstract}


\section{Introduction}\label{intro}
The increasing volume of data has led to the widespread use of cloud computing, which enables users to leverage several computing machines for processing data in a parallel fashion. In order to distribute data among computing machines of a cloud, distributed data processing frameworks are used as an integral part of modern distributed data processing systems. These frameworks distribute the storage and processing of data over a cluster of computing nodes. Hadoop \cite{1}, Spark \cite{2} and Flink \cite{3} are the most widely used frameworks for developing distributed data processing systems \cite{4}. Hadoop is one of the earliest framework and follows the functional programming model of MapReduce. Spark is a novel data processing framework that is designed to overcome the problems faced in Hadoop and Flink is the latest entry into the market that offers features for both batch and stream processing. 

Given the high computational demand, distributed data processing frameworks are deployed and operated via cloud infrastructure \cite{pu2015low}. A cloud can be deployed in three models – private, public, and hybrid. When a cloud is made available to use on a pay-as-you-go basis to the general public, it is called a public cloud. A private cloud is exclusively used and maintained by an organization, not available to the general public. A hybrid cloud is a combination of private and public clouds, offering the best of both worlds \cite{8}. The cloud bursting model of hybrid clouds enables an application to use the private cloud resources and burst into the public cloud when the application faces a shortage on the private side. Hybrid cloud benefits its users in terms of security, cost, resilience and so on \cite{8}. For example, if the data comprises of confidential and non-confidential information, the confidential data is processed via private part and the non-confidential data is burst into the public part of a hybrid cloud \cite{9}. In the meanwhile, organizations can minimize cost by only paying for the temporary resources acquired from the public cloud under spiking workloads instead of paying for purchasing, programming, and maintaining private resources that would remain idle for most of the time \cite{10}.

Several studies have been conducted to evaluate the performance of distributed data processing frameworks in private and public clouds. Mavridis et al. \cite{11} and Dimopoulos et al. \cite{12} have compared the performance of Hadoop and Spark in private clouds. In \cite{14}, the authors have compared Hadoop with Spark deployed in public clouds. Similarly, the performance of Spark and Flink have been contrasted in private clouds \cite{marcu2016spark} as well as public clouds \cite{16}. However, none of the previous studies have comparatively evaluated the performance of Hadoop, Spark and Flink in hybrid clouds. In comparison to private and public clouds that offer computational resource located at one data centre, the impact of distance between the public and private data centres in a hybrid cloud plays a significant part in the performance evaluation of these frameworks. Moreover, although the previous studies have compared Hadoop with Spark or Spark with Flink, there is a paucity of empirical research on apple-to-apple comparison among Hadoop, Spark and Flink. To fill these gaps, this study contributes to the body of knowledge by comparatively evaluating the performance of Hadoop, Spark and Flink in a hybrid cloud. 

In this paper, we report on the implementation of a hybrid cloud consisting of private and public cloud. We then leverage the hybrid cloud to evaluate the performance of the three most widely used frameworks (Hadoop, Spark and Flink). The evaluation is carried out in terms of \textit{execution time}, \textit{resource utilization (CPU, RAM and disk)}, \textit{cost} and \textit{scalability}. Such evaluation aims to determine which hybrid cloud configuration works the best for which distributed data processing framework. It also aims to determine the impact on execution time when more nodes are borrowed by a private cloud from a public cloud. Scalability is a critical concern for data processing systems due to the fluctuating workload \cite{4}. For example, a banking system may experience higher workload at the end of a month to process monthly salary transactions. Consequently, distributed data processing systems should scale up/down based on users' needs. Whilst the \$ cost for private cloud is mostly upfront and relevant to maintenance, the \$ cost for public cloud depends upon the usage especially that of VMs and bandwidth. Therefore, the \$ cost being another key metric in hybrid clouds is considered in our evaluation. 

For the aimed evaluation, we implemented a hybrid cloud consisting of OpenStack\footnote{\label{openstack}https://www.openstack.org/} (the private cloud located in the CREST lab\footnote{https://www.crest-centre.net/}, Adelaide, Australia South region) and Azure\footnote{https://azure.microsoft.com/en-au/} (the public cloud located Sydney, Australia East region). We configured Hadoop, Spark and Flink over different combinations of private and public cloud nodes/Virtual Machines (VM). Thus, we varied the number of VMs and the number of cores in the public cloud to respectively study the horizontal and vertical scalability of the frameworks in a hybrid cloud \cite{el2014scaling}. We also measured the resources (i.e., CPU, RAM, disk and network) utilized during the execution of the experiment. Moreover, we measured the data transfer and data received among the nodes during experimental executions.  
% For studying vertical scalability, we hosted one VM in the public cloud, where the number of cores of the VM changes from 4 to 16. For horizontal scalability, we considered cloud setups where the number of VMs were varied in public cloud from 2 to 8. For the sake of a fair comparison, the allocation of resources (e.g., RAM) between configurations in vertical and horizontal scalability investigation remains the same (see Section \ref{scenarios}). 
% (Xiaoyu: since those information is used to explain horizontal and vertical scalability experiments and also described in the following sections, I think it would be safe to remove them here.)
For this evaluative research study, we used both batch and iterative workloads that are commonly used for evaluating Hadoop, Spark and Flink \cite{18}. In a nutshell, this paper makes the following contributions. 
\begin{enumerate}
    \item We build a hybrid cloud spanning OpenStack and MS Azure for the implementation and evaluation of distributed data processing frameworks in hybrid clouds.
    \item We evaluate and compare the performance of Hadoop, Spark and Flink deployed in the hybrid cloud. Our evaluation reveals several insights in terms of execution time, scalability, data transferred/received, resource utilization and data processing cost of Hadoop, Spark and Flink.
\end{enumerate}

The rest of this paper is organized as follows. Section \ref{frameworks} introduces the studied frameworks. Section \ref{hyrbid-section} reports the hybrid cloud implementation. Section \ref{experimental-setup} describes our experimental setup. Section \ref{results} presents the results, which are followed by practical observations reported in Section \ref{lessons}. Section \ref{related} presents related work and Section \ref{conclusion} concludes the paper.

\section{Distributed Data Processing Frameworks}
\label{frameworks} 
Distributed data processing frameworks are employed in a system for distributing the storage and processing of data across a cluster of nodes. In this study, we have selected the three most popular frameworks, Hadoop, Spark and Flink, based on the fact that they are the most widely used and investigated frameworks in industry and academia \cite{4,19,30}. \textbf{Hadoop} is one of the earliest and widely adopted disk-based data processing framework. As an open-source batch processing framework, it follows the MapReduce functional model by Java \cite{1}. It is composed of two layers - a data storage layer called Hadoop Distributed File System (HDFS) and a data processing layer - Hadoop MapReduce Framework. \textbf{Spark} is a novel framework launched to overcome the problems faced while using Hadoop (e.g., user interface and language compatibility) \cite{marcu2016spark},\cite{20}. Unlike Hadoop, Spark is a memory-based framework with HDFS as its input source and output destination. Before an operation, the user driver program launches multiple workers to read data from a distributed file system and cache them in the memory as partitions of Resilient Distributed Dataset (RDD). This feature enables Spark to avoid reloading data from disk at each iteration and boost the data processing speed. In spark, most of the maps are processed before reduce process starts. \textbf{Flink} is relatively a new open-source memory-based framework suitable for batch and stream processing \cite{3,19}. Fink uses a high throughput streaming engine written in Java and Scala. Similar to Hadoop and Spark, Flink follows a master-slave architecture. However, unlike Spark that implements iterations as for loops, Flink executes iterations as cyclic data flows. The iterative process in Flink significantly speedups certain algorithms by reducing the work in each subsequent iteration. For details on these frameworks, interested readers can refer to \cite{1,2,3,4}.

\section{Hybrid Cloud Implementation}\label{hyrbid-section}
In this section, we describe the implementation of the hybrid cloud used for the evaluation of Hadoop, Spark and Flink.

\subsection{Private and Public Clouds}

Hybrid cloud combines a private cloud with a public cloud. Here, we used OpenStack and Azure as the private and public cloud centers. OpenStack is an open-source framework, which possesses a set of tools to create and manage highly scalable and flexible private clouds. Azure is a software solution for creating, deploying and testing public clouds. To implement the hybrid cloud, we used on-demand usage model (also called cloud bursting) where consumer is a private cloud and donor is a public cloud. In such a setup, under spiking workload, a private cloud borrows resources (e.g. VMs) from a public cloud.

\subsection{Cloud Connectivity}
A secure connection between private and public clouds is a significant part of a hybrid cloud because VMs are deployed across different networks in both clouds \cite{4}. Azure provides various VPN connectivity solutions, however, most of them are costly. To build a secure and cost-free connection between OpenStack and Azure, we opted to use WireGuard \cite{22} which is a Linux kernel-based VPN. WireGuard is user-friendly and offers several benefits such as security, reliability in connection, zero-cost and high throughput \cite{23}.

\begin{figure}
\centering
% \centerline{\includegraphics[width=0.9\linewidth]{figs/data/Hybrid-cloud-new.pdf}}
\centerline{\includegraphics[width=0.8\linewidth]{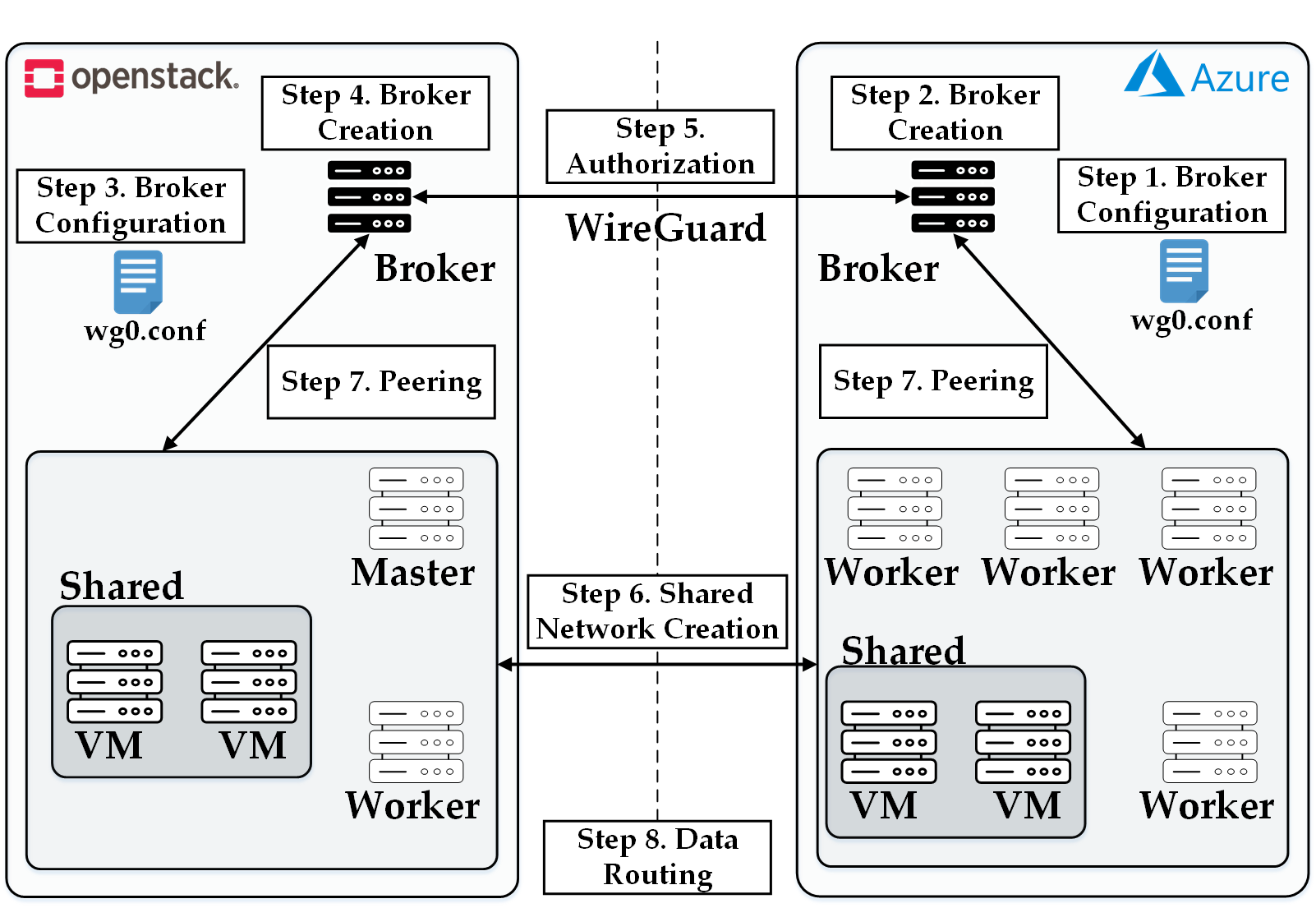}}
\caption{Steps of hybrid cloud implementation}
\vspace{-2 em}
\label{Hybrid-cloud}
\end{figure}

\subsection{Infrastructure Resource Deployment}
\label{infras-deploy}
Several tools, such as command-line interface and portable web services, are available to deploy resources in a public cloud. However, those tools only support deploying limited resources with static behaviours. Therefore, to deploy resources in a dynamic way, we leveraged Terraform \cite{24} – an open-source tool for cloud resource management. Terraform enabled us to define and execute resource deployment in large-scale clusters in a repetitive manner with little user's interference.

\subsection{Hybrid Cloud Implementation}
\label{implement-hybrid}
Here, we outline the steps we followed for setting up a hybrid cloud. The implementation, inspired from \cite{23}, consists of 8 steps as shown in \autoref{Hybrid-cloud}. \textit{Step 1 - Azure broker configuration:} First, we used Wireguard to create private and public keys, which are used to create the configuration file (wg0) of Wireguard for broker VM on Azure. \textit{Step 2 - Azure broker creation:} This step used Terraform to create Azure broker and then install Wireguard on the broker. \textit{Step 3 - OpenStack broker configuration:} Like step 1, this step used Wireguard to create private key and public key but for OpenStack broker. Then, the Wireguard config file (wg0) was created using public keys of OpenStack and Azure broker VMs. \textit{Step 4 - OpenStack broker creation:} This step leveraged Terraform to create OpenStack broker and install Wireguard on the broker VM. \textit{Step 5 - Authorization of OpenStack Broker VM:} In order to connect OpenStack broker VM to the Azure broker VM, the public key of OpenStack broker VM was added to the Wireguard configuration file (wg0) of the Azure broker VM. \textit{Step 6 - Shared networks creation:} This step created shared networks (for hosting VMs) on Azure and OpenStack. \textit{Step 7 - Peering:} Azure broker network was connected to Azure shared network and OpenStack broker network was connected to OpenStack shared network. \textit{Step 8 - Data Routing:} This step implemented data routing in Azure and OpenStack. In Azure, the broker network’s routing table was linked with the shared network. Then, the routing table was updated in accordance with the shared network. In OpenStack, data routing was implemented by simultaneous addition of static rules to the router in the broker and shared networks.

\begin{figure}[!t]
\captionsetup{justification=centering}
\centering
  \begin{subfigure}{0.4\textwidth}
  \centering
  \includegraphics[width=0.9\linewidth]{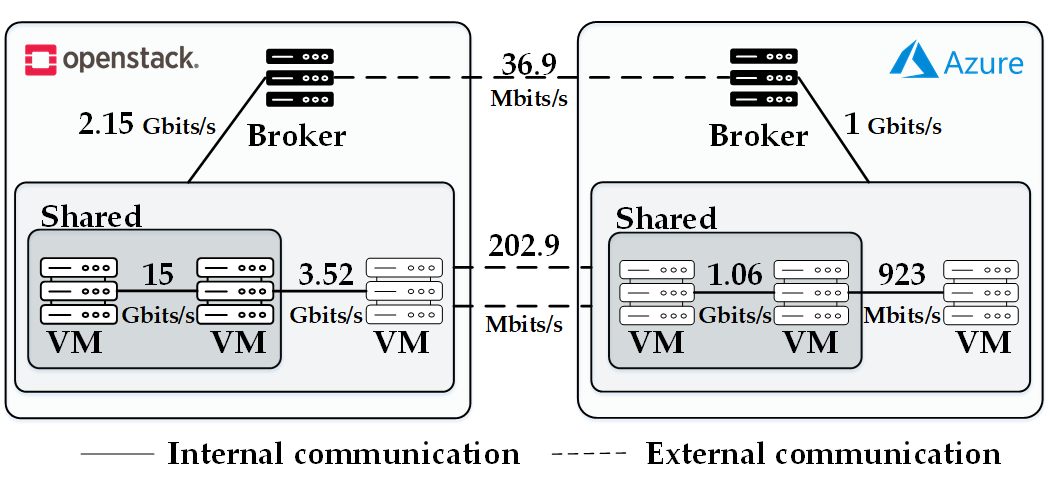}
\end{subfigure}
\vspace{-0.5 em}
\caption{Bandwidth between different pairs of nodes in our hybrid cloud setup}
\label{bandwidth}
\vspace{-0.5 em}
\label{bandwidth-values}
\vspace{-1.0 em}
\end{figure}

After the implementation, a bandwidth measurement experiment was performed in our hybrid cloud. This experiment is important for understanding the results presented in Section \ref{results} as bandwidth among nodes directly impacts the execution time. IPerf3\footnote{https://iperf.fr/}, a cross-framework tool, was installed on all nodes to measure the bandwidth as presented in \autoref{bandwidth}. The network connection between VMs residing on the same physical machine in the OpenStack achieves the mean bandwidth of 15 Gbits/s while it is only 1.06 Gbits/s in Azure. The mean bandwidth between VMs residing on different physical machines is comparatively lower i.e., 3.52 Gbits/s in OpenStack and 923 Mbits/s in Azure. In contrast, the bandwidth of the WAN connecting the VMs in the OpenStack to the VMs in Azure is recorded as 202.91 Mbits/s. Hence, it is evident that the bandwidth between the nodes in the same cloud is much higher than that between a node in OpenStack and a node in Azure. 

\section{Experimental Setup}\label{experimental-setup}
%In this section, we describe the experimental setup used for conducting this study. 

\subsection{Benchmarking Workloads}\label{workloads}

Grep and K-means were selected as benchmarking workloads. Grep is a batch workload that implements one-pass processing, where the input is processed exactly once, for searching plain text. This workload uses Wikipedia text files \cite{marcu2016spark}. K-means is a well-known algorithm for clustering data \cite{18}. It is an iterative workload that implements loop-caching, where the same input is processed multiple times. This workload receives input from the random data generator. We chose data sizes as 5GB for each workload in view of two factors: (i) the time and cost incurred by the range of experiments conducted and (ii) the reliability of the experimental findings. Conducting large-scale experiments on Azure costs a significant amount - 0.03\$/hour for one vCPU and 0.13\$/GB data sent. To ensure cost limitation do not threaten the reliability of our findings, we compared our data size with the state-of-the-art studies (e.g., \cite{14},\cite{zhang2019meteor}) on Hadoop Spark and Flink. We found that the chosen data size is equal to or larger than the most related works such as 350 MB in \cite{14} and 640 MB in \cite{zhang2019meteor}.

\subsection{Infrastructure}
\label{infrastructure}

Our experimental setup consisted of 17 VMs – one master and 16 worker nodes deployed on the hybrid cloud that consists of private (OpenStack) and public (Azure) clouds. Since conducting large-scale experiments on a public cloud incurs a significant cost, we had to choose a reasonable cluster size while avoiding the threats to the reliability of our findings. We first consulted the related studies (e.g., 8 VMs in \cite{14}, 6 VMs in \cite{11},\cite{20}, 4 VMs in \cite{25} and 3 VMs in \cite{16}), and chose a cluster consisting of 17 VMs. Then, we conducted an experiment to observe the trend with the increase in the number of VMs. As presented in Section \ref{cluster-scaling-section}, the trend continues smoothly when the number of VMs increases. Hence, even a further increase in the number of VMs is unlikely to contradict our findings. For evaluating the impact of cloud bursting on the execution time, we considered different combinations of VMs that could be distributed and deployed on private and public clouds, denoted by \textit{(x\_y)}, where \textit{x} and \textit{y} respectively represent the number of VMs in private and public clouds. The configurations are (16\_0), (14\_2), (12\_4), (10\_6), (8\_8), (6\_10), (4\_12) and (2\_14)\footnote{Throughout the text and figures/tables, \textit{(x\_y)} denotes hybrid cloud configuration where \textit{x} and \textit{y} respectively represent the number of VMs deployed in the private and public cloud.}.

\begin{table}[!t]
\captionsetup{justification=centering}
\caption{Specification of the virtual machines}
\vspace{-1.5 em}
\scriptsize
\renewcommand{\arraystretch}{1.1}
\begin{center}
\begin{tabular}{l||c|c}
\hline
% \textbf{Feature}&\multicolumn{3}{|c|}{\textbf{Table Column Head}} \\
\textbf{Feature} \hspace{1 em} &  \hspace{0.5 em} \textbf{Private Cloud (OpenStack)} \hspace{0.5 em} &  \hspace{0.5 em} \textbf{Public Cloud (Azure)} \hspace{0.5 em}\\
% \textbf{} & \textbf{(OpenStack)} & \textbf{(Azure)}\\
\hline
\hline
CPU & 1 vCPU & 1 vCPU\\
%\hline
RAM & 2 GB & 2 GB\\
%\hline
Disk & 10 GB & 30 GB\\
%\hline
Location & Adelaide & Sydney\\
% \textbf{Head} & \textbf{\textit{Table column subhead}}& \textbf{\textit{Subhead}}& \textbf{\textit{Subhead}} \\
% \hline
% copy& More table copy$^{\mathrm{a}}$& &  \\
\hline
% \multicolumn{4}{l}{$^{\mathrm{a}}$Sample of a Table footnote.}
\end{tabular}
\vspace{-3 em}
\label{summaryofvms}
\end{center}
\end{table}

In our setup, the master VM, equipped with 2 core CPU, 4GB RAM and 40 GB disk, is hosted in the private cloud, while the distribution of worker VMs varies based on cloud configurations. The specification of the worker VMs used in OpenStack and Azure is presented in \autoref{summaryofvms}. We use Terraform \cite{24} for reliably creating and destroying VMs in the private and public clouds. 
Given that our study required frequent creation and destruction of VMs, configuration of distributed data processing frameworks and installation of benchmarking suites, we have automated the whole process via bash scripts to ensure the minimal human interference. During the experiments, we always destroy the used cluster and create a fresh one to avoid the impacts of previous run/setup. Each experiment is executed three times to remove abruptness and the mean results are reported.

\subsection{Experimental Scenarios}\label{experimental-scenarios}
\label{scenarios}
The experimental scenario underpins the way experiment is executed. In our scenario, the user connects with the master node deployed in the private cloud using SSH connection. Using the master node, \textit{step 1} is to generate the dataset as per the workload using benchmarking suites such as HiBench \cite{17} and BigDataBench\cite{18}. In \textit{step 2}, the dataset is uploaded to HDFS from the local file system of the nodes deployed on private and public clouds. In \textit{step 3}, the measurement probes for various metrics presented in Section \ref{metrics} are activated. In \textit{step 4}, the data processing job starts. Once data processing is completed, the results are recorded into a file. The execution time reported in Section \ref{results} is the time consumed in \textit{step 4}, i.e., data processing. The time consumed in data generation, data transmission, and uploading/downloading data to/from HDFS is reported separately in Section \ref{results_cloud_bursting}. In this study, the data processing time is the focus to evaluate the three frameworks since these frameworks operate independently of data generation, data transfer and data upload/download.

\subsection{Evaluation Metrics}\label{metrics}
During our experiments, we measured 10 metrics i.e., execution time, data transferred, data received, vertical scalability, horizontal scalability, cost, CPU usage, RAM usage, disk read and disk write. 
% We measured execution time through time counter that starts at the beginning of data processing and ends as soon as the data processing is completed. 
The time consumed in each data processing phase (e.g, map and reduce) is calculated based on the log file analysis. We used iftop\footnote{https://www.tecmint.com/iftop-linux-network-bandwidth-monitoring-tool/}, running on each node to measure the data transferred and received during the experiment execution. Besides, Dstat\footnote{https://linux.die.net/man/1/dstat} is used to measure resource utilization (i.e., CPU, RAM, and disk). The horizontal scalability is measured by increasing the number of nodes in the public cloud, and the vertical scalability is measured by increasing the number of cores of a single node in the public cloud (details in Section \ref{scalability-s}). 

% \begin{figure}[t]
% \centerline{\includegraphics{figs/checking.jpg}}
% \caption{Example of a figure caption.}
% \label{fig}
% \end{figure}

\begin{figure}
\captionsetup{justification=centering}
\centering
\begin{subfigure}{.21\textwidth}
  %\centering
  \includegraphics[width=\linewidth]{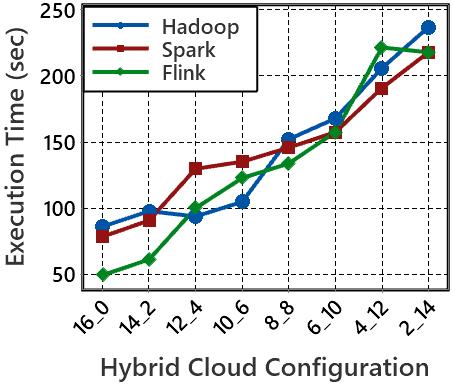}
  \vspace{-1.5 em}
  \caption{Grep}
  %\vspace{5.00mm}
  \label{cloud-burst-a}
\end{subfigure}
\hspace{1 em}
\begin{subfigure}{.21\textwidth}
  %\centering
  \includegraphics[width=\linewidth]{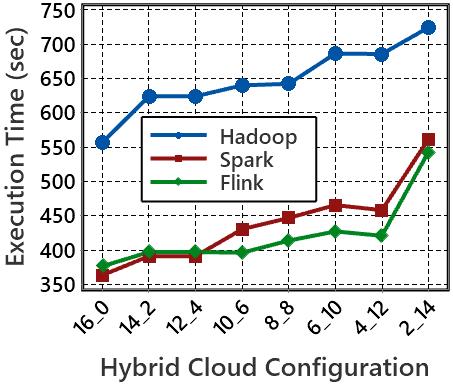}
  \vspace{-1.5 em}
  \caption{K-means}
  %\vspace{5.00mm}
  \label{cloud-burst-b}
\end{subfigure}%
\vspace{-0.5 em}
\caption{Execution time of Hadoop, Spark and Flink for Grep and K-means with various hybrid cloud configurations 
% specified by x-y, where x and y denotes the number of nodes deployed on private and public cloud
} 
\vspace{-0.5 em}
\label{bursting_grep_kmeans}
\end{figure}

\begin{figure}
\captionsetup{justification=centering}
\centering
\vspace{-0.3 em}
\begin{subfigure}{.35\textwidth}
  %\centering
  \includegraphics[width=\linewidth]{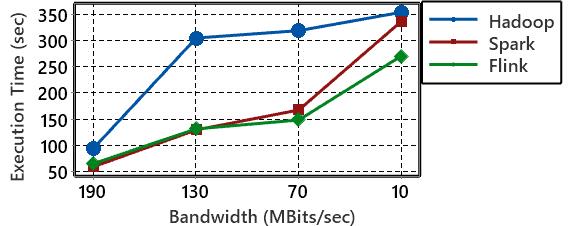}
  %\vspace{5.00mm}
\end{subfigure}
\vspace{-0.7 em}
\caption{Impact of bandwidth of the network connection between private and public cloud for (8\_8) configuration}
\label{bandwidth-impact}
\vspace{-2 em}
\end{figure}

\begin{figure*}
\captionsetup{justification=centering}
\centering
\begin{subfigure}{.16\textwidth}
  \centering
  \includegraphics[width=\linewidth]{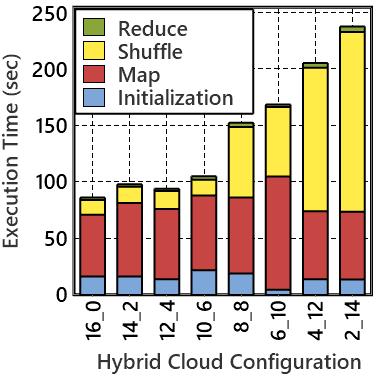}
  \caption{Hadoop-Grep}
  %\vspace{5.00mm}
  \label{phases-a}
\end{subfigure}%
\hspace{0.2 em}
\begin{subfigure}{.16\textwidth}
  \centering
  \includegraphics[width=\linewidth]{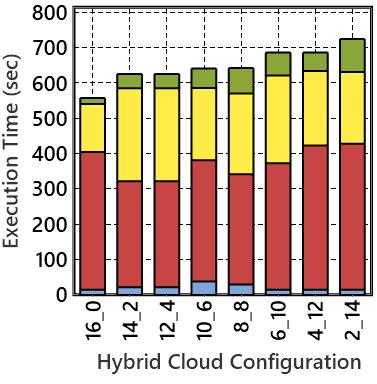}
  \caption{Hadoop-Kmeans}
  %\vspace{5.00mm}
  \label{phases-b}
\end{subfigure}%
\hspace{0.2 em}
\begin{subfigure}{.16\textwidth}
  \centering
  \includegraphics[width=\linewidth]{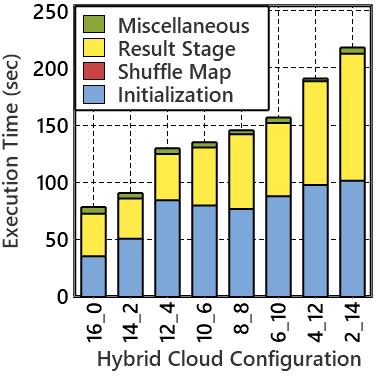}
  \caption{Spark-Grep}
  %\vspace{5.00mm}
  \label{phases-c}
\end{subfigure}
\begin{subfigure}{.16\textwidth}
  \centering
  \includegraphics[width=\linewidth]{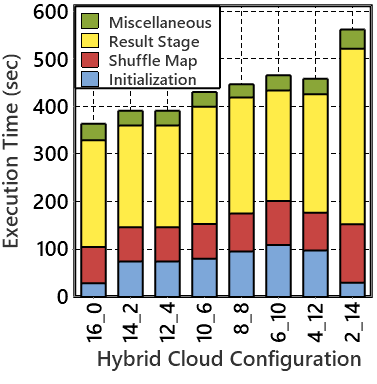}
  \caption{Spark-Kmeans}
  %\vspace{5.00mm}
  \label{phases-d}
\end{subfigure}
\begin{subfigure}{.16\textwidth}
  \centering
  \includegraphics[width=\linewidth]{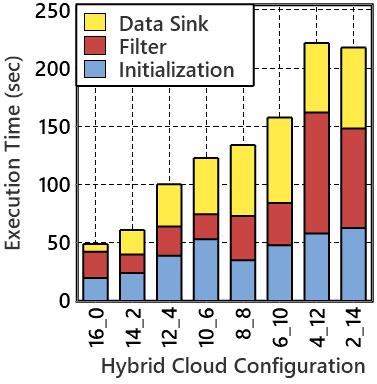}
  \caption{Flink-Grep}
  %\vspace{5.00mm}
  \label{phases-e}
\end{subfigure}
\begin{subfigure}{.16\textwidth}
  \centering
  \includegraphics[width=\linewidth]{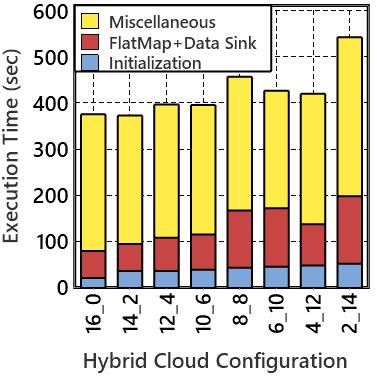}
  \caption{Flink-Kmeans}
  %\vspace{5.00mm}
  \label{phases-f}
\end{subfigure}
\caption{Execution time per stages of the data processing. Same legend applies to Fig (a) and (b)}
% In n\_m at x-axis, n and m denote the number of VMs deployed on the private and public cloud respectively
\label{framework-phases}\vspace{-2.5 em}
\hspace{2.0 em}
\end{figure*} 

\section{Results}\label{results}
%This section presents our detailed results. 
\subsection{\textit{Impact of Cloud Bursting}}\label{results_cloud_bursting}
 Fig. \ref{bursting_grep_kmeans} shows the execution time of Hadoop, Spark and Flink in the hybrid cloud for batch (Grep) and iterative (K-means) workloads. Generally, for all three frameworks, the execution time increases when more nodes are borrowed by the private cloud from the public cloud, i.e., as we move from non-bursting (16\_0) to full-bursting (2\_14). This can be attributed to two factors - network bandwidth and disk I/O. As depicted in Fig. \ref{bandwidth-impact}, the bandwidth between nodes in the private cloud is 3.52 Gbits/s, between nodes in the public cloud is 1.06 Gbits/s, and across private and public cloud (WAN) is 202 Mbits/s. As we borrow more nodes from the public cloud, more connections with lower bandwidth are used. This is evident from the amount of data transferred between nodes in various hybrid cloud configurations illustrated in Fig. \ref{data-transfer} (details in Section \ref{data-transfer-received}). To illustrate the impact of bandwidth, we ran a small-scale experiment where we used Traffic Control (a linux utility) to decrease the bandwidth of the network connection between private and public cloud for (8\_8) configuration (executing grep) by 32\% of the original value. The result depicted in Fig. \ref{bandwidth-impact} reveals that as the bandwidth decreases, the execution time increases. As more nodes are exploited in the public cloud, the disk usage increases as depicted in Fig. \ref{resource-utilization-maps} (details in Section \ref{resource-utilization}). This is because the nodes in the public cloud are equipped with 30 GB disk as compared to 10 GB disk on nodes in the private cloud, as shown in Table \ref{summaryofvms}. % For instance, in 12\_4 configuration, the disk read rate is 1.8 MB/sec, which increases to 2.3 MB/sec in 4\_12 configuration. Similarly, the disk write rate increases from 1.5 MB/sec to 2.3 MB/sec as we move from 14\_2 to 2\_14 configuration. 

\begin{figure*}
\captionsetup{justification=centering}
\centering
\begin{subfigure}{.3\textwidth}
  \centering
  \includegraphics[width=\linewidth]{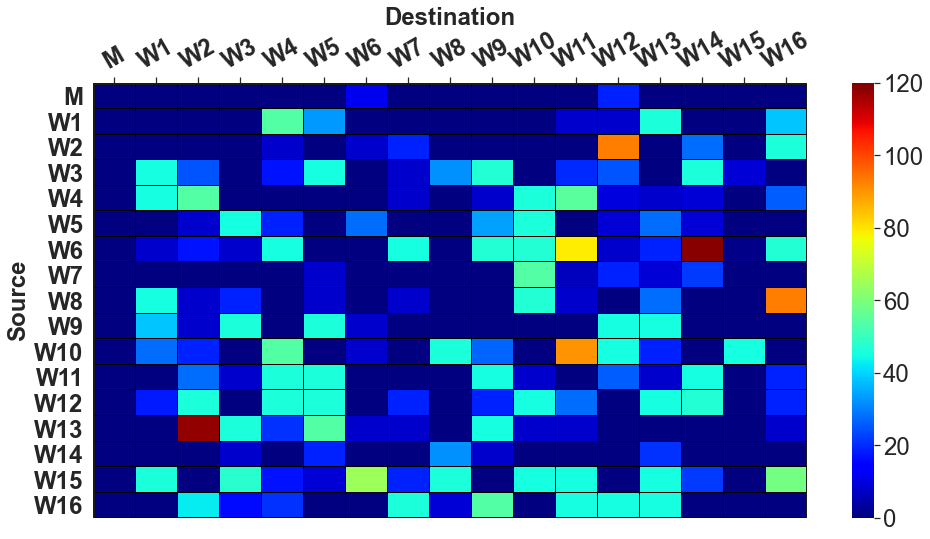}
  \caption{Hadoop}
  %\vspace{5.00mm}
  \label{vertical-d}
\end{subfigure}%
\hspace{1 em}
\begin{subfigure}{.3\textwidth}
  \centering
  \includegraphics[width=\linewidth]{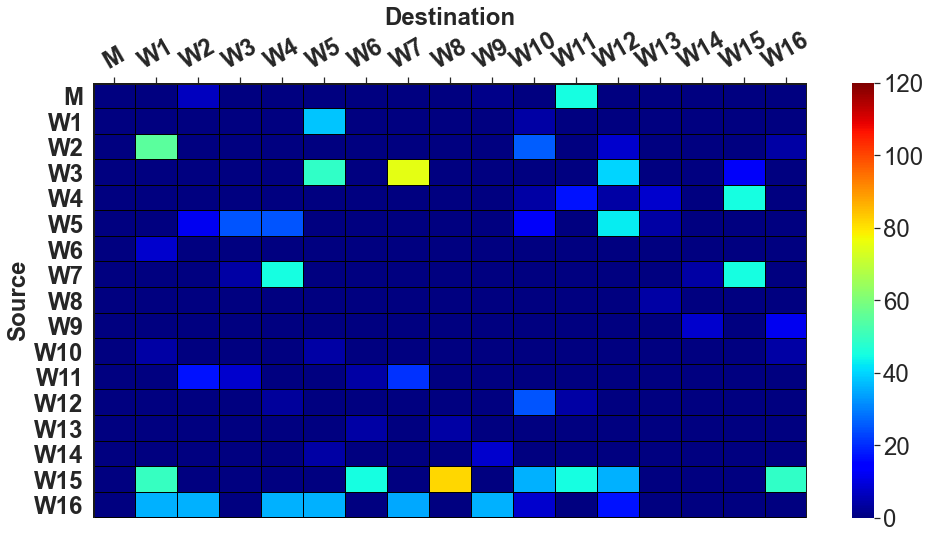}
  \caption{Spark}
%  %\vspace{5.00mm}
  \label{vertical-e}
\end{subfigure}%
\hspace{1 em}
\begin{subfigure}{.3\textwidth}
  \centering
  \includegraphics[width=\linewidth]{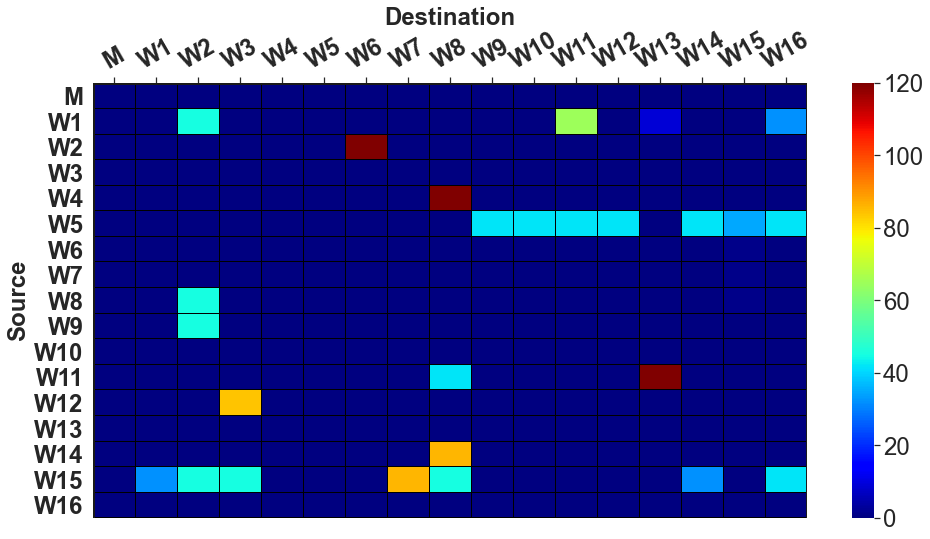}
  \caption{Flink}
  %\vspace{5.00mm}
  \label{vertical-f}
\end{subfigure}%
\hspace{1 em}

\caption{Data transfer/received (in MB) among the nodes executing K-means in (8\_8). M denotes master node and W1-W16 denotes the worker nodes.}
% X-axis specifies the number of cores (i.e., 4, 8, and 16) of the VM deployed in public cloud center
\label{data-transfer-maps}
\vspace{-1 em}
\end{figure*}

\begin{figure*}
\captionsetup{justification=centering}
\centering
\begin{subfigure}{.225\textwidth}
  \centering
  \includegraphics[width=\linewidth]{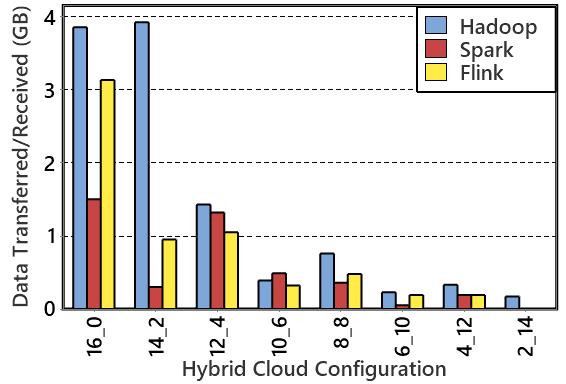}
  \caption{Within Private Cloud}
  %\vspace{5.00mm}
  \label{data-transfer-a}
\end{subfigure}%
\hspace{0.5 em}
\begin{subfigure}{.225\textwidth}
  \centering
  \includegraphics[width=\linewidth]{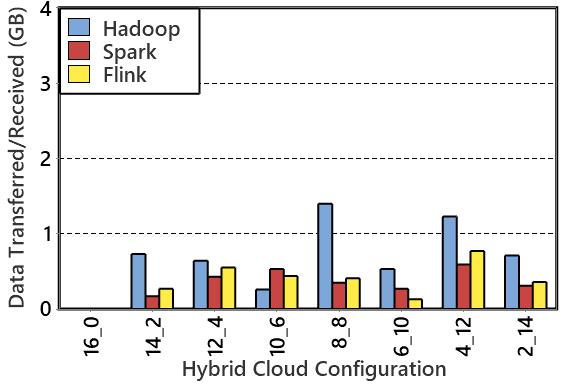}
  \caption{Private to Public Cloud}
  %\vspace{5.00mm}
  \label{data-transfer-b}
\end{subfigure}%
\hspace{0.5 em}
\begin{subfigure}{.225\textwidth}
  \centering
  \includegraphics[width=\linewidth]{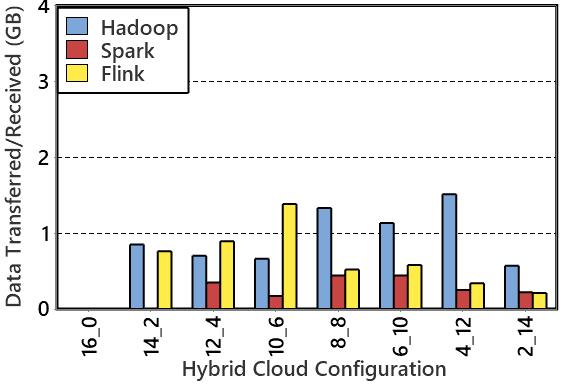}
  \caption{Public to Private Cloud}
  %\vspace{5.00mm}
  \label{data-transfer-c}
\end{subfigure}
\hspace{0.5 em}
\begin{subfigure}{.225\textwidth}
  \centering
  \includegraphics[width=\linewidth]{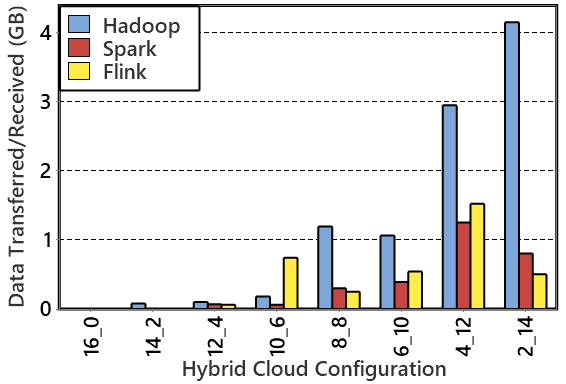}
  \caption{Within Public Cloud}
  %\vspace{5.00mm}
  \label{data-transfer-d}
\end{subfigure}
\label{data-transfer}\vspace{-0.5 em}
\caption{Data transferred/received in our hybrid cloud infrastructure during the execution of K-means}
% In n\_m at x-axis, n and m denote the number of VMs deployed on the private and public cloud respectively
\label{data-transfer}\vspace{-1.8 em}
\end{figure*}

%This is because of the difference in the data communication pattern and bandwidth among the nodes in the various hybrid cloud configurations. In order to understand this, Figure ?? depicts three hybrid cloud configurations - one with most node deployment on the private cloud, one with most node deployment in the public cloud, and one with the equal node deployment in public and private cloud. Each of the three hybrid cloud configuration has three type of network connections labeled as C\_High, C\_Medium, and C\_Low in Figure ??. The connection among the nodes within the private cloud, labelled as C\_High, has the highest bandwidth i.e., 26.5 Mbits/s. The connection among the nodes within the public cloud, labeled as C\_Medium, has a bandwidth of 14.4 Mbits/s while the C\_Low, which is the connection between the nodes in the private and public cloud has lowest bandwidth of 8.1 Mbits/s. As presented in Figure ??, for 14\_2, most of the data transfer occurs within the nodes among the private cloud using C\_High, which results in a lower execution time. For 8\_8, the most used connection type is C\_Low, which is why the execution time increases. Finally, for 2\_14, most data transfer happens across the two clouds between the nodes in the private and the nodes in the public cloud. Whilst this this data transfer happens through C\_Medium, it also partly leverages C\_High, hence, combining the usage of a medium and low-speed connection, which results in a further increase in the execution time. 

On average, Hadoop is around 27.5\% and 33.6\% slower than Spark and Flink, respectively. This is attributed to the disk-based architecture of Hadoop that, unlike Spark and Flink significantly leverages the disk during data processing.
% As shown in Section \ref{resource-utilization}, Hadoop respectively writes 2.3 MB to the disk per second. Comparatively, Spark and Flink write only 1.19 MB and 1.13 MB data to the disk respectively.
Since disk-based operations are computationally heavy and time-consuming, it slows Hadoop down. This trend continues across almost all cloud configurations as shown in Fig. \ref{resource-d}. Furthermore, Hadoop spends a lot of time in data shuffling stage during which data is written to the disk in a sequential manner. Flink outclasses Spark by around 8.1\% in terms of execution time. Whilst both Spark and Flink are memory-based frameworks, the in-built optimizer of Flink enables it to use CPU and RAM more efficiently as reported in Section \ref{resource-utilization}. When moving from non-bursting (16\_0) to full-bursting (2\_14), the mean execution time of Hadoop, Spark and Flink increases by 3.4$\times$, 2.2$\times$ and 2.1$\times$, respectively. The slowdown in Hadoop is higher as compared to Spark and Flink because Hadoop transfers more data among the nodes during data processing as shown in Fig. \ref{total-data-transfer}. For instance, Hadoop, Spark and Flink transfer around 4 GB, 2.5 GB and 2.7GB data during the execution of K-means. 
%This data transfer occurs via network especially WAN, which slows Hadoop down as the usage of the network increases with increase in the number of nodes exploited in the public cloud. 
With respect to workloads, the difference among Hadoop, Spark and Flink is not as evident for batch workload as compared to iterative workloads. This is because Hadoop is originally designed for batch workload while Spark and Flink are designed for iterative workloads. Unlike Hadoop, Spark and Flink cache intermediate data across nodes between iterations to improve efficiency. Fig. \ref{bursting_grep_kmeans} also shows that the mean execution time respectively increases by 3.4$\times$ and 1.4$\times$ for batch and iterative workloads as we move from (16\_0) to (2\_14).  

In Fig. \ref{bursting_grep_kmeans}, the trend of increase in execution time is not exactly smooth when borrowing more nodes from the public cloud. This is a consequence of two uncontrolled factors - the fluctuation in WAN bandwidth and load distribution among workers. Except (16\_0), all the hybrid cloud configurations leverage the WAN connection for data transfer between private and public clouds. When the strength of the WAN connection fluctuates, it directly impacts the execution time. Secondly, the load distribution among the workers is not even by default, which further impacts the execution time. As shown in Fig. \ref{resource-utilization-maps}, worker W12 leverages only 41\% CPU and 0.42GB RAM with Spark during the execution of K-means while other workers are used to a higher extent (49 - 55\%). The under utilization of workers may be caused by multiple reasons including the network congestion and software failure. As an example, a worker may not be available to respond to the master. Then, the master node has to send more workloads to other available workers. As mentioned in Fig. \ref{bursting_grep_kmeans} and Fig. \ref{bandwidth-impact}, the execution time here is the time consumed by the frameworks during data processing. The time consumed in data generation, uploading/downloading data to/from HDFS/local file system, and data transfer is presented in Table \ref{data-generation-time}. Similar to data processing time, the time for other operations is also higher when more nodes are deployed in the public cloud. In addition, more time is consumed in uploading data from the local file system to HDFS on public cloud. 
%Whilst such distribution of load can be controlled in these frameworks via parameter tuning, we did not consider it in this study as we used the default configuration of these frameworks.

\begin{table}
\captionsetup{justification=centering}
\caption{Time (sec) consumed during various operations in the three hybrid cloud configurations. 
%In (x\_y), x and y respectively denote the number of nodes deployed on the private and public cloud.
}
\centering
\vspace{-0.5 em}
%\scriptsize
\begin{tabular}{l||r|r|r}
\hline
\textbf{Operation} & \textbf{14\_2} & \textbf{8\_8} & \textbf{2\_14} \\ \hline \hline
Data generation & 104 & 240 & 415 \\ 
Downloading data from HDFS to local & 60 & 177 & 299 \\ 
Data transfer from private to public cloud & 529 & 375 & 932 \\ 
Uploading data to HDFS from local & 902 & 1,106 & 1,352 \\
\hline
\end{tabular}
\vspace{-2 em}
\label{data-generation-time}
\end{table}

\subsection{Data Processing Phases}
Fig. \ref{framework-phases} shows the time consumed in each phase of data processing. It is evident that Spark takes significant time (around 73s) in framework initialization. This is because when the execution of a Spark job starts, it creates SparkContext which is a slow process. 
% This time can be decreased through parameter tuning or reusing SparkContext, which is outside the scope of this work. 
Flink takes around 45s in initialization. Contrary to Spark and Flink, Hadoop takes minimal time (around 5s) to initialize. These results imply that the actual data processing time is not dominated by the framework initialization time in our study. As expected, the framework initialization time is similar for the executions of grep and k-means. However, the phases for the execution of grep and k-means are different in Spark and Flink. Fig. \ref{framework-phases} shows that the time consumed in each phase is negligibly impacted by the hybrid cloud configuration. Spark and Flink do not have a shuffle map phase for Grep because Spark and Flink only perform the filter transformation and result stage in the execution of Grep \cite{marcu2016spark}. 
% Similarly, the phases for executing Grep and K-means with Flink are different from each other. Also, several phases run in parallel with each for both Spark and Flink. Hence, it is not possible to present them separately. In Fig. \ref{framework-phases}, these phases are represent as one phase termed as miscellaneous. The phases included in miscellaneous for Flink include map, reduce, bulk iterations, combine reduce, flatmap collect and sync. 
Comparing Fig. \ref{phases-b} and Fig. \ref{phases-d}, we observe that Hadoop takes longer time in shuffle phase as compared to Spark. This is because during the shuffle stage, Hadoop workers write data to the disk in a sequential manner facing synchronization barrier, where each thread has to wait for the antecedent thread to finish writing. On the other hand, Spark cache most of the data in memory during shuffle phase since Spark is a memory-based framework having no synchronization barrier.

\subsection{Data Transfer/Received}\label{data-transfer-received}
Fig. \ref{data-transfer-maps} shows the data transferred between each pair of the nodes during the execution of K-means by Hadoop, Spark and Flink under (8\_8) hybrid cloud configuration. We can observe that Hadoop engages almost all nodes in large data transfer. This is because Hadoop always transfers the whole data among all nodes while Spark and Flink can reduce the size of data in memory after serialization \cite{shi2015clash}. In Spark and Flink, only a few nodes receive large data. For example, with Spark, only worker W15 made one large size data (85.2MB) transfer to worker W8. The master node communicates uneven data ranging from 0.03 - 19.9MB to workers in Hadoop and 0.16 - 45MB in Spark, while even size data ranging from 0.03 - 0.12MB are transferred from master to workers in Flink. As the data transferred from the workers to the master is concerned, all three frameworks transfer almost even data ranging from 0.27 - 0.73MB to the master node. We did not observe a significant difference in data transfer with respect to hybrid cloud configurations (e.g., (16\_0) and (2\_14)). Therefore, we only report the data transfer among the nodes for one representative configuration i.e., (8\_8). 

\begin{figure}
\captionsetup{justification=centering}
\centering
\begin{subfigure}{.28\textwidth}
  %\centering
  \includegraphics[width=\linewidth]{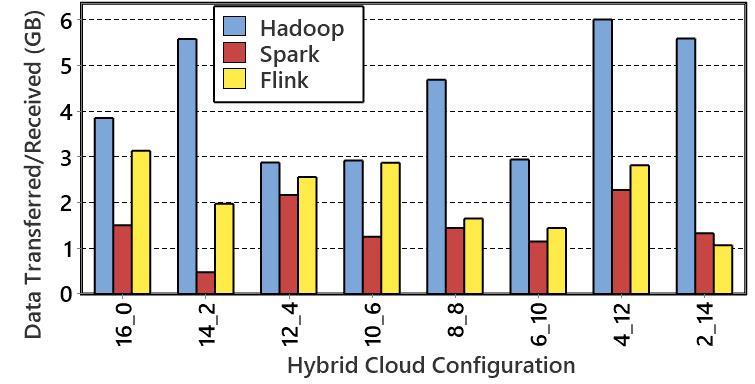}
  %\vspace{5.00mm}
\end{subfigure}
\vspace{-0.5 em}
\caption{Total data transferred/received among the nodes during the execution of K-means.}
\label{total-data-transfer}
\vspace{-1 em}
\end{figure}

Fig. \ref{total-data-transfer} shows the total amount of data communicated among the nodes during the execution of K-means. Overall, Hadoop transfers 4.31 GB of data among the nodes, which is followed by Flink (2.19 GB) and Spark (1.45 GB). The data transfer for Spark is quite low because transformations on RDD in Spark are lazy in nature, which helps Spark to minimize data transfer between the nodes. Unlike the total data transfer shown in Fig. \ref{total-data-transfer}, Fig. \ref{data-transfer} depicts the data transfer in or between public and private clouds. For (16\_0), all data transfers happened among the nodes within the private cloud, since there are no nodes deployed in the public cloud. For (2\_14), most of the data is transferred among the nodes in the public cloud as 14 out of 16 nodes are deployed in the public cloud. For (8\_8), most of the data transfers happened across the clouds using WAN. These results show that these data processing frameworks do not consider the underlying cloud infrastructure (e.g., bandwidth) to optimize the data transfer among the nodes.

\begin{figure*}
\captionsetup{justification=centering}
\centering
\vspace{-1 em}
\begin{subfigure}{.40\textwidth}
  \centering
  \includegraphics[width=\linewidth]{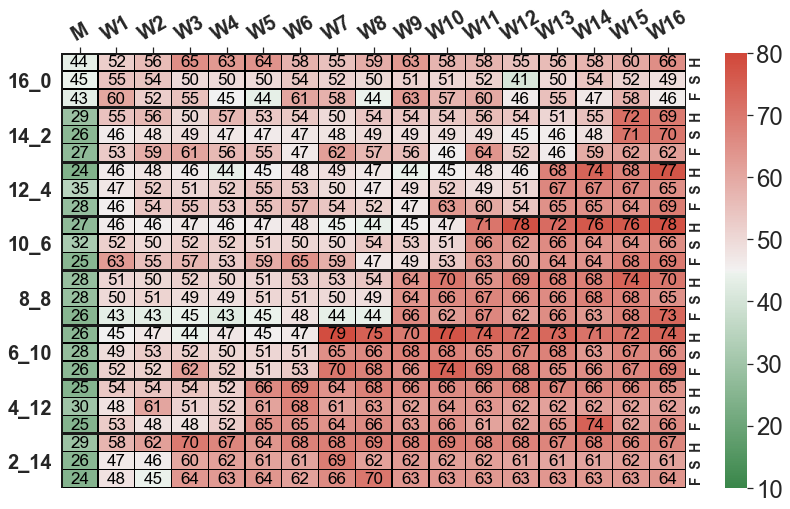}
%   \vspace{-0.5 em}
  \caption{CPU Usage (\%)}
  %\vspace{5.00mm}
  \label{resource-a}
\end{subfigure}%
\hspace{1 em}
\begin{subfigure}{.40\textwidth}
  \centering
  \includegraphics[width=\linewidth]{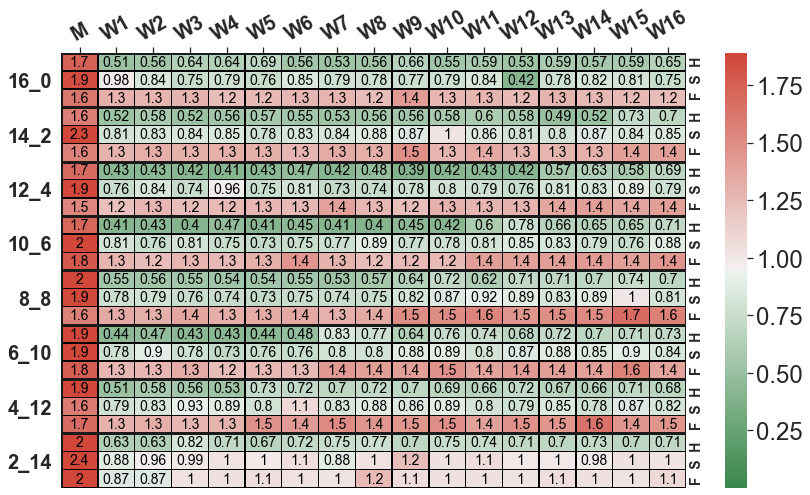}
  \caption{RAM Usage (GB)}
  %\vspace{5.00mm}
  \label{resource-b}
\end{subfigure}%
\hspace{1 em}
\begin{subfigure}{.40\textwidth}
 \centering
  \includegraphics[width=\linewidth]{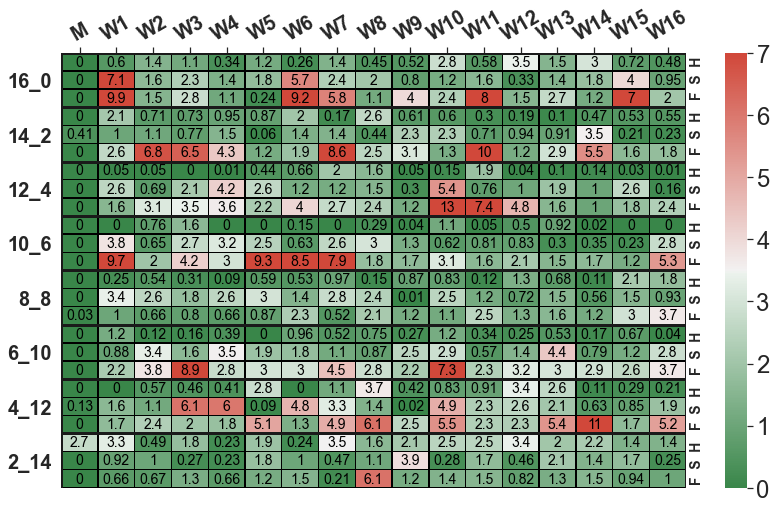}
  \caption{Disk Read (MB)}
  %\vspace{5.00mm}
  \label{resource-c}
\end{subfigure}%
\hspace{1 em}
\begin{subfigure}{.40\textwidth}
  \centering
  \includegraphics[width=\linewidth]{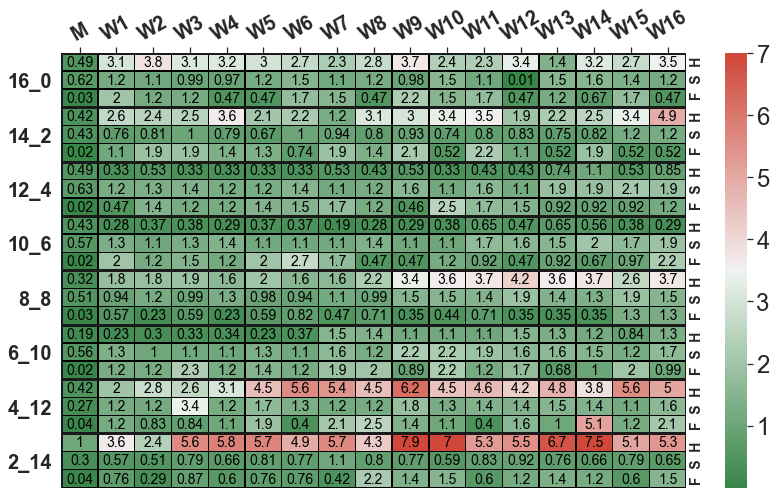}
  \caption{Disk Write (MB)}
  %\vspace{5.00mm}
  \label{resource-d}
\end{subfigure}%
\hspace{1 em}
\vspace{-0.5 em}
\caption{Resource utilization during the execution of K-means. H, S, and F denotes Hadoop, Spark and Flink, respectively. 
% M denotes the master node and W1-W16 denotes the 16 worker nodes. In x\_y, x and y respectively denotes the number of nodes in the private and public cloud
}
\label{resource-utilization-maps}
\vspace{-1 em}
\end{figure*}

\subsection{Resource Utilization}\label{resource-utilization}

Fig. \ref{resource-utilization-maps} shows the mean utilization of CPU, RAM and disk during the execution of K-means. In each hybrid cloud configuration shown in Fig. \ref{resource-utilization-maps}, the first set of nodes is deployed in the private cloud and the last set in the public cloud. For example, in (14\_2), the workers W1 - W14 are deployed in the private cloud and W15-W16 are deployed in the public cloud. The values presented in Fig. \ref{resource-utilization-maps} are averaged over per second values collected using Dstat. We observe that the CPU usage of the nodes in the public cloud is higher as compared to the nodes in the private cloud. This is because the CPU capacity of the nodes in the public cloud is 2.4 GHz as compared to the 3.0 GHz CPU capacity in the private cloud. Therefore, the nodes in public cloud utilize CPU more to catch up with the private cloud nodes. The CPU usage of master node is quite low compared to the worker nodes. This is because the master node acts as a manager only without directly contributing to the job execution. Similarly, the percentage use of RAM by master node is lower than that by worker nodes. However, the RAM usage of the master node is higher than that of worker nodes in Fig. \ref{resource-b} because the master node is equipped with 4 GB RAM while worker nodes are equipped with 2 GB RAM. Overall, the CPU usage ranges from 24-79\%, which reveals two facts (i) all the nodes worked well in our cluster; (ii) the CPU usage did not reach 100\% due to the job dependency during data processing. For example, W1 needs a particular piece of data to execute its task. However, this piece of data is produced by W2. Therefore, W1 has to wait for W2 until the required data is produced, and this results in reducing the overall CPU usage. As expected, the RAM usages of Spark and Flink are higher than that of Hadoop. However, Flink's RAM usage is also higher than Spark. A potential reason for this is that Spark leverages JVM's heap memory, while Flink maintains its memory stack, which is more optimally designed.  

Fig. \ref{resource-c} and Fig. \ref{resource-d} depict the disk read and disk write. The disk read and write for master node is much lower than the worker nodes. On average, Hadoop, Spark and Flink write 2.38 MB/s, 1.19 MB/s and 1.13 MB/s to the disk. This is why Hadoop is slower than Spark and Flink since disk I/O operations are computationally heavy. In terms of disk read, the trend is opposite where Hadoop, Spark and Flink read around 0.85 MB/s, 1.69 MB/s and 3.00 MB/s from the disk, respectively. We also observe that the disk writes for nodes deployed in the public cloud are higher than nodes in the private cloud. A potential reason is the higher disk availability (30 GB) of nodes in the public cloud, while nodes in the private cloud only have 10 GB disk.

\subsection{Horizontal and Vertical Scalability}\label{scalability-s}
Horizontal scalability refers to scaling-up a system by adding new computing nodes, while vertical scalability refers to the scale-up of a system by adding more cores into the same computing node. Whilst previous studies (e.g., \cite{4},\cite{14,marcu2016spark,16}) have explored the scalability of these frameworks in private or public clouds, we investigate how these frameworks scale in the hybrid cloud. For assessing horizontal scalability, we fixed one VM of default settings (1 vCPU and 2GB RAM) in the private cloud and vary the number of default VMs in the public cloud from 1 to 2 and 2 to 4. For vertical scalability investigation, we deployed one default VM in the private cloud and one default VM in the public cloud. Then, we varied the number of cores of the VM deployed in the public cloud from 1 to 2 and 2 to 4. The single VM in private cloud is deployed in order to adhere to the definition of the hybrid cloud. For the sake of fair comparison, same resources (e.g., RAM size) are allocated to the various setups in vertical and horizontal scalability. For example, a VM with 4 cores in vertical scalability investigation and 4 VMs in horizontal scalability investigation are equipped with 8 GB RAM each.  

\begin{figure}
\captionsetup{justification=centering}
\centering
\begin{subfigure}{.2\textwidth}
  %\centering
  \includegraphics[width=\linewidth]{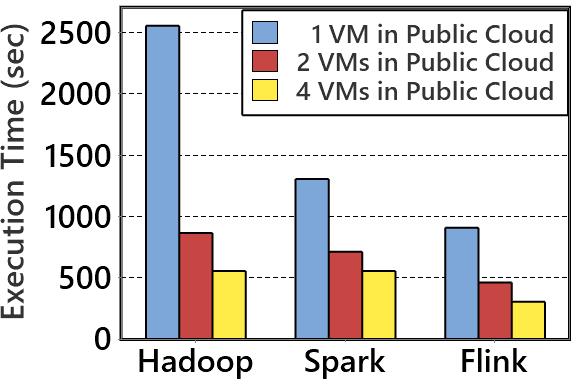}
  %\vspace{5.00mm}
  \caption{Horizontal scalability}
  \label{horizontal}
\end{subfigure}
\hspace{1 em}
\begin{subfigure}{.2\textwidth}
  %\centering
  \includegraphics[width=\linewidth]{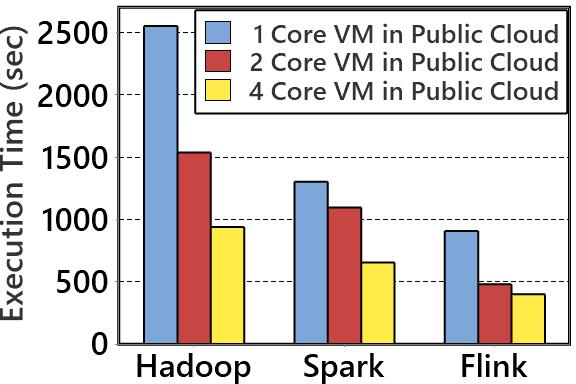}
  %\vspace{5.00mm}
  \caption{Vertical scalability}
  \label{vertical}
\end{subfigure}
\vspace{-0.5 em}
\caption{Scalability as the number of VMs (horizontal) or number of cores (vertical) increases in the public cloud}
\label{scalability}
\vspace{-0.5 em}
\end{figure}

\begin{figure}
\captionsetup{justification=centering}
\centering
\begin{subfigure}{.23\textwidth}
  %\centering
  \includegraphics[width=\linewidth]{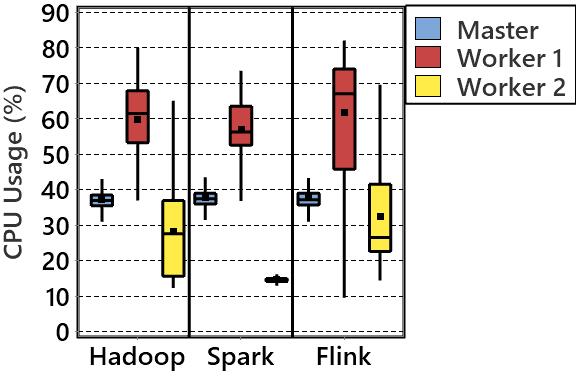}
  %\vspace{5.00mm}
  \caption{3 Node Cluster}
  \label{cpu-scalability-3}
\end{subfigure}
%\hspace{1 em}
\begin{subfigure}{.23\textwidth}
  %\centering
  \includegraphics[width=\linewidth]{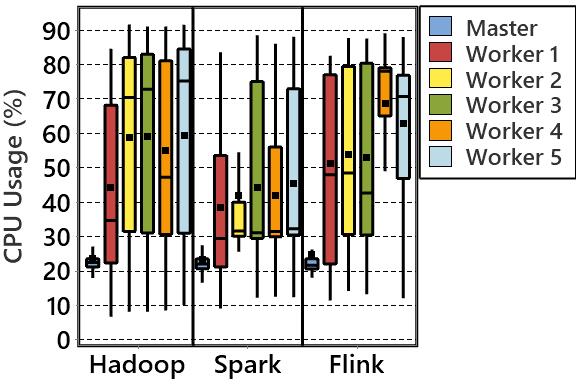}
  %\vspace{5.00mm}
  \caption{6 Node Cluster}
  \label{cpu-scalability-5}
\end{subfigure}
\vspace{-0.5 em}
\caption{CPU Usage in cluster with 3 nodes (vertical scaling) and with 6 nodes (horizontal scaling)}
\label{cpu-scalability}
\vspace{-2 em}
\end{figure}

Fig. \ref{scalability} shows the impacts of horizontal and vertical scalability on Hadoop, Spark and Flink in the hybrid cloud. As expected, the execution time decreases with the increases in the number of VMs (horizontal scalability) or the number of cores (vertical scalability). When doubling the number of VMs, the mean execution time reduces by 51.0\%, 33.7\% and 41.7\% for Hadoop, Spark and Flink, respectively. When doubling the number of cores, the execution time is reduced by 39.3\%, 28.1\% and 31.9\% for Hadoop, Spark and Flink, respectively. This shows that horizontal scalability leads to more reduction in execution time compared to vertical scalability in a hybrid cloud. This is because in vertical scalability, the scale-up is bottle-necked by the low-capacity node in the private cloud due to non-optimal load distribution. Therefore, the increase in the capacity of the node in the public cloud is not fully utilized. As shown in Fig. \ref{cpu-scalability-3}, worker 1 is of 2 cores while worker 2 is of 8 cores. Therefore, worker 2 is only utilized around 25\% since it is dependent upon a weaker node i.e., worker 1. In Fig. \ref{cpu-scalability-5}, all workers are of 2 cores, hence, all workers are almost equally utilized.  

\subsection{Cluster Scaling}\label{cluster-scaling-section}
\begin{figure}
\captionsetup{justification=centering}
\centering
\begin{subfigure}{.35\textwidth}
  %\centering
  \includegraphics[width=\linewidth]{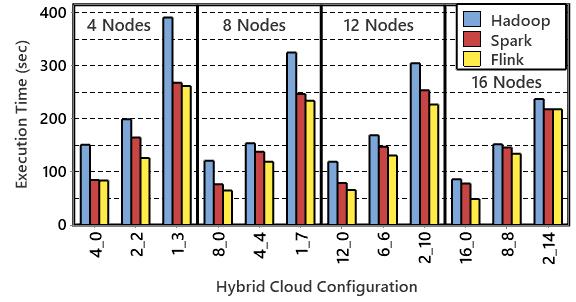}
  %\vspace{5.00mm}
\end{subfigure}
\vspace{-0.5 em}
\caption{Execution time of grep in 4, 8, 12 and 16 node cluster. 
}
\label{cluster-scaling}
\vspace{-1.5 em}
\end{figure}

Unlike the previous section where we scale the cluster only in the public cloud, in this section, we scale the overall cluster size. The motivation for such scaling is to determine the trend with respect to the increasing number of nodes. Moreover, this experiment justifies the cluster size (16 nodes)  used for the experiments reported in this paper. We conducted the experiment with various numbers of nodes, including 4, 8, 12 and 16 nodes, when executing Grep. As shown in Fig. \ref{cluster-scaling}, with the increase in the number of nodes, the execution times of all the three frameworks decrease because of the increase in the hybrid cloud capacity. In addition, when moving from non-bursting, i.e., (4\_0), (8\_0), (12\_0) and (16\_0) to full-bursting, i.e., (1\_3), (1\_7), (2\_10) and (2\_14), the execution times of Hadoop, Spark and Flink also increase. This is same as the analysis in Section \ref{results_cloud_bursting}. Moreover, Hadoop always consumes the highest execution time, followed by Spark and then Flink. Therefore, we conclude that the various configurations could impact the performance of these frameworks in the experiments. However, they do not impact the performance comparison among these frameworks in general. Thus, the experiment results in Section \ref{results} obtained using our experimental setup reported in Section \ref{infrastructure} are valid.

%\textcolor{blue}{In the later phase of the execution, Flink becomes inefficient in terms of resource usage due to its filter-> count operator. Since grep involves several filter operations, Flink becomes affected unlike Spark which takes advantages of its persistant control over the RDDs (Spark vs Flink paper)}

\begin{figure}
\captionsetup{justification=centering}
\centering
\begin{subfigure}{.32\textwidth}
  %\centering
  \includegraphics[width=\linewidth]{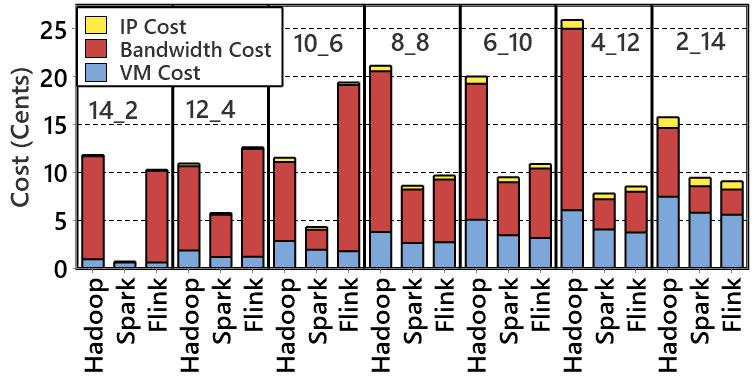}
  %\vspace{5.00mm}
\end{subfigure}
\vspace{-0.5 em}
\caption{Cost incurred on MS Azure for K-means execution
% In (x\_y), x and y respectively denote the number of VMs deployed on OpenStack and Azure.
}
\vspace{-2 em}
\label{cost}
\end{figure}

\subsection{Cost Analysis}
The cost incurred by the frameworks deployed in various cloud configurations is calculated using Equation \ref{equation1}, which underpins the costs of VM, bandwidth and IP. Those three aspects account for around 95\% of the total cost in MS Azure. In Equation (\ref{equation1}), $V_{N}$ denotes the number of VMs, \textit{T} denotes the time (in hours) VM and associated IP are used for, $D_{S}$ denotes the data (in GB) sent out from the public cloud, $I_{N}$ denotes the number of IPs, \textit{C} denotes the hybrid cloud configuration, \textit{P} denotes the framework, and \textit{W} denotes the workload. Furthermore, 0.0264, 0.126 and 0.004 are respectively the rates (in USD) for VM, bandwidth and IP in MS Azure (Australia East region). 
\begin{equation}
\small 
\vspace{-0.5 em}
\begin{split}
    Cost (C, P, W) = & (V_N\times T\times 0.0264) + (D_S\times 0.126) \\
    & + (I_N\times T\times 0.004)
\end{split}
\vspace{-0.5 em}
\label{equation1}
\end{equation}
The costs incurred by each framework and hybrid cloud configuration during the execution of K-means are presented in Fig. \ref{cost}. Since the number of IPs used is the same for the three frameworks, the main cost determinants are VM time and bandwidth for sending data. In majority of the cases, Hadoop is the most expensive followed by Flink, while Spark is the least expensive. In (12\_4) and (10\_6) configuration, Flink is slightly more expensive than Hadoop, which is attributed to the large data transfers (i.e., 0.89 GB and 1.38 GB) as compared to 0.70 GB and 0.66 GB data transfer by Hadoop as illustrated in Fig. \ref{data-transfer-c}. Hadoop is more expensive because Hadoop's execution time is higher as well as Hadoop transfers significantly larger data among nodes. However, it is worth noting that execution time and cost are only two quality attributes. There are several other quality attributes (e.g., fault tolerance and usability) according to which Hadoop outclasses Spark and Flink \cite{4}. Moreover, Hadoop is a well-established framework that is in use for years in industry, hence, its replacement with the recent frameworks is not a straight-forward undertaking \cite{4}. Fig. \ref{cost} also shows that unlike expected, the cost does not increase consistently as more and more nodes are borrowed from the public cloud. This is because in addition to the number of VMs, the bandwidth used contributes significantly to the overall cost. For instance, the mean cost of VM, bandwidth and IP is 3.14 cents, 7.98 cents, and 0.47 cents, respectively.

\begin{figure}
\captionsetup{justification=centering}
\centering
\begin{subfigure}{.3\textwidth}
  %\centering
  \includegraphics[width=\linewidth]{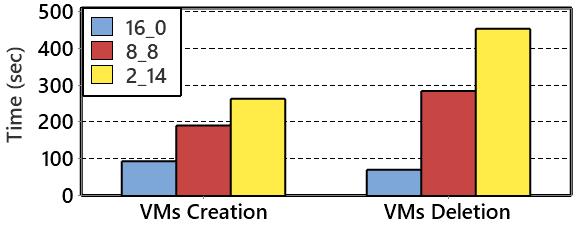}
  %\vspace{5.00mm}
\end{subfigure}
\vspace{-0.7 em}
\caption{Time consumed in creating and destroying VMs}
\label{vm-creation-deletion}
\vspace{-1.8 em}
\end{figure}

\section{Practical Observations and Future Work}
\label{lessons}
In this section, we present our practical observations acquired during this study and directions for future research.

\textbf{Infrastructure deployment in Azure is slow:} Our hybrid cloud infrastructure requires dynamically creating and destroying VMs in both Azure and OpenStack using Terraform. However, such operations in Azure take much longer time than OpenStack. Fig. \ref{vm-creation-deletion} shows the time consumed in creating and destroying VMs in three hybrid cloud configurations. When more VMs are utilized in Azure, the time increases. The potential reasons include but not limited to hardware resource allocations from the available pool, uploading OS to VMs and informing the load balancer and firewall of the Azure data center about the new deployments.

\textbf{Automation is the key:} We observed that the hybrid cloud implementation, framework configurations and deployments are complex and tedious. Therefore, we assert that the whole process should be as automated as possible. This not only saves a practitioner/researcher’s time but also ensures the reliability of data processing and associated experiments. In this regard, we recommend the use of Terraform \cite{24} for automatically creating, deploying and destroying resources on a hybrid cloud. Similarly, writing shell scripts for automating the framework configuration is a fruitful practice.

\textbf{Confirm version compatibility:} In this study, we used various frameworks, workloads and software tools in collaboration with each other. Before dwelling into implementation, it is important to confirm that the versions of tools are compatible with each other. For example, we have noticed that Flink 0.10 is only compatible with BigDataBench 5.0, while BigDataBench 5.0 requires Hadoop 2.7. Moreover, Spark 2.4 is compatible with HiBench 7.1 but not with HiBench 7.0, and Hadoop 2.7 is only compatible with Java version 1.8 or above. In addition, Spark and Flink require version compatibility with Hadoop to use HDFS and YARN as the cluster manager. Therefore, we assert that version compatibility should be confirmed before starting the implementation.

\textbf{Future Work:} \textit{Optimizing data transfers:} As shown in Fig. \ref{data-transfer-maps} and Fig. \ref{data-transfer}, the frameworks transfer data randomly among nodes in the private cloud, public cloud, and between private and public clouds. In comparison to data transfers within the same cloud, the data transfers between two clouds via WAN are costly both in terms of time and money. Therefore, the frameworks should be tuned/configured in a way that minimizes inter-cloud data transfers in a hybrid cloud setup. Whilst each framework comes with over 150 parameters, we suggest focusing on parameters (e.g., data replication factor and number of parallel threads for transferring data) related to data transfer, data distribution and network usage to optimize data transfers especially across the two clouds. \textit{VM hosting:} In this study, the VMs in private and public clouds are hosted either on the same or different physical machines. As presented in Fig. \ref{bandwidth-values}, the bandwidth between VMs hosted on the same physical machine is higher than VMs on different machines. Furthermore, Fig. \ref{bandwidth-impact} also reveals that the bandwidth among VMs directly impacts the execution time. Therefore, it will be fruitful to extend this study for different combinations of VMs placement strategies such as hosting all VMs on the same physical machine, hosting only VMs in the private cloud on the same physical machine, and so on.

\section{Related Work}\label{related}
In this section, we compare our work with the existing studies to position the novelty of our work.

\textbf{Hybrid cloud implementation and cloud bursting:} Several studies (e.g., \cite{9},\cite{23},\cite{33,35,36}) have explored the implementation of the hybrid cloud and cloud bursting. Mansouri et al. \cite{23} used WireGuard \cite{22} to connect private and public clouds for implementing a hybrid cloud to compare the impact of cloud bursting on the performance of six distributed databases (e.g., Cassandra and MongoDB). Clemente-Castelló et al. \cite{35} leveraged a low throughput and high latency link to connect two clouds hosted on OpenStack for building a collaborative hybrid cloud. The authors then used this hybrid cloud for building a performance model to predict the execution time of Hadoop jobs. Similarly, Loreti et al. \cite{9} connected two OpenStack clouds to build a hybrid cloud for evaluating Hadoop performance with respect to bandwidth. The results show that the performance gain via bursting into a public cloud is largely dependent upon the inter-cloud bandwidth. Along the same lines, Roman et al. \cite{36} studied the impact of inter-cloud bandwidth on Spark performance in a hybrid cloud. It was observed that the shuffle phase of Spark is very sensitive to the bandwidth between the two networks (private and public) in the hybrid cloud. Bicer et al. \cite{33} studied the impact of cloud bursting and scalability for data-intensive applications (e.g., KNN) in a hybrid cloud comprising of local and AWS clusters. Taking inspirations from \cite{23}, we have also used WireGuard \cite{22} to implement a hybrid cloud that integrates private cloud (OpenStack) with public cloud (Azure).

\textbf{Comparison of distributed data processing frameworks:} Several studies (e.g., \cite{11},\cite{14},\cite{marcu2016spark},\cite{16},\cite{20},\cite{25}) have evaluated and compared the performance of distributed data processing frameworks. Veiga et al. \cite{20} have used multiple benchmarking workloads to compare the performance of Hadoop, Spark and Flink deployed on a private cloud. The authors concluded that Spark and Flink outperforms Hadoop with an average reduction of 77\% and 70\% in execution time. Gu et al. \cite{14} compared the performance of Hadoop and Spark deployed on a cluster of eight nodes. The findings revealed that Spark outperformed Hadoop with respect to execution time. However, Spark consumed far more memory than Hadoop. Perera et al. \cite{16} compared Spark and Flink. Both the frameworks were deployed on Amazon EC2 instances. Due to pipelined execution, Flink consistently outperformed Spark with respect to execution time. Shi et al. \cite{25} also compared the execution time of Hadoop and Spark. They found that Spark outperformed Hadoop by 2.5× for WordCount, and 5× for K-means and PageRank. In \cite{11}, the performance of Hadoop was compared with Spark by analyzing log files with respect to execution time, resource usage and scalability in a cluster of six VMs on a private cloud. The results showed that by maximizing the resource utilization, Spark outperformed Hadoop, while horizontal scale-up of worker nodes led to reducing the execution time by about 50\%. In \cite{marcu2016spark}, the experiments were executed in private cloud to evaluate Spark and Flink for various workloads. The evaluation showed that Flink performed better for batch workloads whereas Spark performed well in large-scale graph processing.

In comparison to existing studies, this paper fills the following two gaps (i) none of the previous studies have evaluated/compared the execution time, data transfer/received, horizontal scalability, vertical scalability, cost and resource utilization for Hadoop, Spark and Flink in a hybrid cloud, and (ii) none of the previous studies have evaluated the performance of Hadoop, Spark and Flink with different permutations of VMs in the cloud bursting model. Hence, the previous studies are largely orthogonal to this study.

\section{Conclusion}\label{conclusion}
In this paper, we first reported on the implementation of a hybrid cloud consisting of private (OpenStack) and public (MS Azure) clouds. We then used the hybrid cloud for evaluating and comparing the three most widely used distributed data processing frameworks (i.e. Hadoop, Spark and Flink) in terms of execution time, resource utilization, horizontal scalability, vertical scalability and cost. We used both batch and iterative workloads in our evaluation, and found that the execution time of the three frameworks increase as more nodes are borrowed from the public cloud. In a hybrid cloud setup, Flink is the fastest followed by Spark and then Hadoop. With respect to cost, Spark is the least expensive while Hadoop is found the most expensive. All three frameworks horizontally scale better as compared to vertical scaling. In the future, we plan to evaluate the impact of distance between private and public clouds on the performance of distributed data processing frameworks in a hybrid cloud setup.

\bibliographystyle{IEEEtran}
\vspace{-0.5 em}
\bibliography{IEEEabrv,SHagun.bib}

\begin{thebibliography}{10}
\providecommand{\url}[1]{#1}
\csname url@samestyle\endcsname
\providecommand{\newblock}{\relax}
\providecommand{\bibinfo}[2]{#2}
\providecommand{\BIBentrySTDinterwordspacing}{\spaceskip=0pt\relax}
\providecommand{\BIBentryALTinterwordstretchfactor}{4}
\providecommand{\BIBentryALTinterwordspacing}{\spaceskip=\fontdimen2\font plus
\BIBentryALTinterwordstretchfactor\fontdimen3\font minus
  \fontdimen4\font\relax}
\providecommand{\BIBforeignlanguage}[2]{{%
\expandafter\ifx\csname l@#1\endcsname\relax
\typeout{** WARNING: IEEEtran.bst: No hyphenation pattern has been}%
\typeout{** loaded for the language `#1'. Using the pattern for}%
\typeout{** the default language instead.}%
\else
\language=\csname l@#1\endcsname
\fi
#2}}
\providecommand{\BIBdecl}{\relax}
\BIBdecl

\bibitem{1}
Apache, ``Hadoop: An open-source software for reliable, scalable, distributed
  computnig. available at https://hadoop.apache.org/ [last accessed: 23 may
  2021],'' 2009.

\bibitem{2}
M.~Zaharia, R.~S. Xin, P.~Wendell, T.~Das, M.~Armbrust, A.~Dave, X.~Meng,
  J.~Rosen, S.~Venkataraman, and M.~J. Franklin, ``Apache spark: a unified
  engine for big data processing,'' \emph{Communications of the ACM}, vol.~59,
  no.~11, pp. 56--65, 2016.

\bibitem{3}
P.~Carbone, A.~Katsifodimos, S.~Ewen, V.~Markl, S.~Haridi, and K.~Tzoumas,
  ``Apache flink: Stream and batch processing in a single engine,''
  \emph{Bulletin of the IEEE Computer Society Technical Committee on Data
  Engineering}, vol.~36, no.~4, 2015.

\bibitem{4}
D.~Singh and C.~K. Reddy, ``A survey on platforms for big data analytics,''
  \emph{Journal of big data}, vol.~2, no.~1, p.~8, 2015.

\bibitem{pu2015low}
Q.~Pu, G.~Ananthanarayanan, P.~Bodik, S.~Kandula, A.~Akella, P.~Bahl, and
  I.~Stoica, ``Low latency geo-distributed data analytics,'' \emph{ACM SIGCOMM
  Computer Communication Review}, vol.~45, no.~4, 2015.

\bibitem{8}
B.~P. Rimal, E.~Choi, and I.~Lumb, ``A taxonomy and survey of cloud computing
  systems,'' in \emph{International Joint Conference on INC, IMS and
  IDC}.\hskip 1em plus 0.5em minus 0.4em\relax Ieee, 2009, pp. 44--51.

\bibitem{9}
D.~Loreti and A.~Ciampolini, ``A hybrid cloud infrastructure for big data
  applications,'' in \emph{IEEE 17th International Conference on High
  Performance Computing and Communications}.\hskip 1em plus 0.5em minus
  0.4em\relax IEEE, 2015.

\bibitem{10}
A.~N. Toosi, R.~N. Calheiros, and R.~Buyya, ``Interconnected cloud computing
  environments: Challenges, taxonomy, and survey,'' \emph{ACM Computing Surveys
  (CSUR)}, vol.~47, no.~1, pp. 1--47, 2014.

\bibitem{11}
I.~Mavridis and H.~Karatza, ``Performance evaluation of cloud-based log file
  analysis with apache hadoop and apache spark,'' \emph{Journal of Systems and
  Software}, vol. 125, pp. 133--151, 2017.

\bibitem{12}
S.~Dimopoulos, C.~Krintz, and R.~Wolski, ``Big data framework interference in
  restricted private cloud settings,'' in \emph{2016 IEEE International
  Conference on Big Data (Big Data)}.\hskip 1em plus 0.5em minus 0.4em\relax
  IEEE, 2016, pp. 335--340.

\bibitem{14}
L.~Gu and H.~Li, ``Memory or time: Performance evaluation for iterative
  operation on hadoop and spark,'' in \emph{IEEE 10th International Conference
  on High Performance Computing and Communications}.\hskip 1em plus 0.5em minus
  0.4em\relax IEEE, 2013.

\bibitem{marcu2016spark}
O.-C. Marcu, A.~Costan, G.~Antoniu, and M.~S. P{\'e}rez-Hern{\'a}ndez, ``Spark
  versus flink: Understanding performance in big data analytics frameworks,''
  in \emph{2016 IEEE International Conference on Cluster Computing
  (CLUSTER)}.\hskip 1em plus 0.5em minus 0.4em\relax IEEE, 2016, pp. 433--442.

\bibitem{16}
S.~Perera, A.~Perera, and K.~Hakimzadeh, ``Reproducible experiments for
  comparing apache flink and apache spark on public clouds,'' \emph{arXiv
  preprint arXiv:1610.04493}, 2016.

\bibitem{el2014scaling}
I.~El-Helw, R.~Hofman, and H.~E. Bal, ``Scaling mapreduce vertically and
  horizontally,'' in \emph{SC'14: Proceedings of the International Conference
  for High Performance Computing, Networking, Storage and Analysis}.\hskip 1em
  plus 0.5em minus 0.4em\relax IEEE, 2014, pp. 525--535.

\bibitem{18}
S.~Huang, J.~Huang, Y.~Liu, L.~Yi, and J.~Dai, ``Hibench: A representative and
  comprehensive hadoop benchmark suite,'' in \emph{ICDE Workshops}, 2010.

\bibitem{19}
H.~Ahmed, M.~A. Ismail, M.~F. Hyder, S.~M. Sheraz, and N.~Fouq, ``Performance
  comparison of spark clusters configured conventionally and a cloud service,''
  \emph{Procedia Computer Science}, 2016.

\bibitem{30}
W.~Inoubli, S.~Aridhi, H.~Mezni, M.~Maddouri, and E.~M. Nguifo, ``An
  experimental survey on big data frameworks,'' \emph{Future Generation
  Computer Systems}, vol.~86, pp. 546--564, 2018.

\bibitem{20}
J.~Veiga, R.~R. Expósito, X.~C. Pardo, G.~L. Taboada, and J.~Tourifio,
  ``Performance evaluation of big data frameworks for large-scale data
  analytics,'' in \emph{IEEE International Conference on Big Data}, 2016.

\bibitem{22}
J.~A. Donenfeld, ``Wireguard: Next generation kernel network tunnel,'' in
  \emph{NDSS}, 2017.

\bibitem{23}
Y.~Mansouri, V.~Prokhorenko, and M.~A. Babar, ``An automated implementation of
  hybrid cloud for performance evaluation of distributed databases,''
  \emph{Journal of Network and Computer Applications}, 2020.

\bibitem{24}
Y.~Brikman, \emph{Terraform: Up \& Running: Writing Infrastructure as
  Code}.\hskip 1em plus 0.5em minus 0.4em\relax O'Reilly Media, 2019.

\bibitem{zhang2019meteor}
H.~Zhang, H.~Huang, and L.~Wang, ``Meteor: Optimizing spark-on-yarn for short
  applications,'' \emph{Future Generation Computer Systems}, 2019.

\bibitem{25}
J.~Shi, Y.~Qiu, U.~F. Minhas, L.~Jiao, C.~Wang, B.~Reinwald, and F.~Özcan,
  ``Clash of the titans: Mapreduce vs. spark for large scale data analytics,''
  \emph{Proceedings of the VLDB Endowment}, 2015.

\bibitem{17}
L.~Wang, J.~Zhan, C.~Luo, Y.~Zhu, Q.~Yang, Y.~He, W.~Gao, Z.~Jia, Y.~Shi, and
  S.~Zhang, ``Bigdatabench: A big data benchmark suite from internet
  services,'' in \emph{IEEE 20th international symposium on high performance
  computer architecture (HPCA)}.\hskip 1em plus 0.5em minus 0.4em\relax IEEE,
  2014, pp. 488--499.

\bibitem{shi2015clash}
J.~Shi, Y.~Qiu, U.~F. Minhas, L.~Jiao, C.~Wang, B.~Reinwald, and F.~{\"O}zcan,
  ``Clash of the titans: Mapreduce vs. spark for large scale data analytics,''
  \emph{Proceedings of the VLDB Endowment}, 2015.

\bibitem{33}
T.~Bicer, D.~Chiu, and G.~Agrawal, ``A framework for data-intensive computing
  with cloud bursting,'' in \emph{IEEE international conference on cluster
  computing}.\hskip 1em plus 0.5em minus 0.4em\relax IEEE, 2011, pp. 169--177.

\bibitem{35}
F.~J. Clemente-Castelló, B.~Nicolae, R.~Mayo, and J.~C. Fernández,
  ``Performance model of mapreduce iterative applications for hybrid cloud
  bursting,'' \emph{IEEE Transactions on Parallel and Distributed Systems},
  vol.~29, no.~8, pp. 1794--1807, 2018.

\bibitem{36}
R.-I. Roman, B.~Nicolae, A.~Costan, and G.~Antoniu, ``Understanding spark
  performance in hybrid and multi-site clouds,'' 2015.

\end{thebibliography}

\end{document}